\documentclass[nofootinbib,floatfix,superscriptaddress,prd,twocolumn]{revtex4}
\usepackage{graphicx}
\usepackage{epsfig}
\usepackage{bm}
\usepackage{comment}
\usepackage[T1]{fontenc}
\usepackage[latin9]{inputenc}
\usepackage{amssymb}
\usepackage{float}
\usepackage{amsmath}
\usepackage{dcolumn}
\usepackage{cancel}
\usepackage[colorlinks]{hyperref}
\usepackage[usenames,dvipsnames]{color}
\hypersetup{
     breaklinks=true,
    pdfstartview={FitH},    % fits the width of the page to the window
    colorlinks=true,       % false: boxed links; true: colored links
    linkcolor=blue,          % color of internal links
    citecolor=red,        % color of links to bibliography
    filecolor=magenta,      % color of file links
    urlcolor=blue,           % color of external links
    anchorcolor=green,      % Color for anchor text
    linktocpage=true
}

\newcommand{\be}{\begin{equation}}
\newcommand{\ee}{\end{equation}}

\begin{document}

%%%%%%%%%%%%%%%%%%%%%%%%%%%%%%%%%%%%%%%%%%%%%%%%%%%%%%%%%%%%%%%%%%%%%%
\title{Generalized Uncertainty Principle mimicking dynamical Dark Energy: matter perturbations and gravitational wave data analysis}
%
%Imprint of quantum deformed entropy in primordial gravitational waves and Ice cube
%high-energy neutrinos}
%%%%%%%%%%%%%%%%%%%%%%%%%%%%%%%%%%%%%%%%%%%%%%%%%%%%%%%%%%%%%%%%%%%%%%%%%%%%%%%%%%%%%%

\author{G.~G.~Luciano}
\email{giuseppegaetano.luciano@udl.cat }
\affiliation{Department of Chemistry, Physics and Environmental and Soil Sciences, Escola Politecninca Superior, Universidad de Lleida, Av. Jaume
II, 69, 25001 Lleida, Spain}

\author{Y.~Sekhmani}
\email{yassine_sekhmani@um5.ac.ma}
\affiliation{Center for Theoretical Physics, Khazar University, 41 Mehseti Street, Baku, AZ1096, Azerbaijan.}
\affiliation{Centre for Research Impact \& Outcome, Chitkara University Institute of Engineering and Technology, Chitkara University, Rajpura, 140401, Punjab, India.}

\date{\today}

\begin{abstract}
The Generalized Uncertainty Principle (GUP) stands out as a nearly ubiquitous feature in quantum gravity modeling, predicting the emergence of a minimum length at the Planck scale. Recently, it has been shown to modify the area-law scaling of the Bekenstein-Hawking entropy, giving rise to deformed Friedmann equations within Jacobson's approach. The ensuing model incorporates the GUP correction as a quintessence-like dark energy that supplements the cosmological constant, influencing the dynamics of the early Universe while aligning with the $\Lambda$CDM paradigm in the current epoch. 
In this extended scenario, we examine the growth of matter perturbations and structure formation employing the Top-Hat Spherical Collapse approach. Our analysis reveals that the profile of the density contrast is sensitive to the GUP parameter $\beta$, resulting in a slower 
gravitational evolution of primordial fluctuations in the matter density.
%formation of large-scale structures compared to the standard case. 
We also discuss implications for the relic density of Primordial Gravitational Waves (PGWs), identifying the parameter space that enhances the PGW spectrum. Using the sensitivity of the next-generation GW observatories in the frequency range below $10^3\,\mathrm{Hz}$, we constrain $\beta\lesssim10^{39}$, which is more stringent  than most other cosmological/astrophysical limits. This finding highlights the potential role of GWs in the pursuit of understanding quantum gravity phenomenology.
 \end{abstract}

 \maketitle

\section{Introduction}
\label{Intro} 
The $\Lambda$CDM ($\Lambda$ + cold dark matter) paradigm represents the cornerstone of modern cosmology due to its efficacy in explaining the evolutionary history of the Universe. In spite of its numerous virtues, several challenges have been emerging in the past few years at both theoretical and observational levels~\cite{Perivolaropoulos:2021jda}. 
Foremost among these is the discovery of the late-time accelerating expansion of the cosmos~\cite{SupernovaSearchTeam:1998fmf,SupernovaCosmologyProject:1997czu}, which has prompted extensive inquiries aimed at establishing its origin. The original hypothesis involves invoking the cosmological constant $\Lambda$, which appears as a manifestation of the vacuum energy. To align with observations, $\Lambda$ should be much smaller than the Planck scale $l_p$, i.e., $\Lambda\simeq10^{-122}\,l_p^{-2}$.  
Nevertheless, the theorized vacuum energy  alarmingly exceeds this value, giving rise to the  cosmological constant tension\footnote{We use units where $\hslash=c=K_B=1$.}~\cite{Weinberg:1988cp}. 

An alternative scheme explores dynamical dark energy (DE) models~\cite{Peebles:1987ek,Sahni:1999gb,Copeland:2006wr}. In this context, the possibility of a scalar field with self-interaction potential density has been examined~\cite{Caldwell:1997ii,Coble:1996te}, following the observation that any candidate theory of relativistic gravity should include a matter-coupled scalar field for consistency with the Mach principle~\cite{Brans:1961sx}. 
Notably, observational evidence appears to favor dynamical DE models over the $\Lambda$CDM~\cite{Zhao:2017cud}, stimulating intensive effort to go beyond the standard scenario. 

In the challenge posed by the dark side of the Universe, valuable insights could be provided by quantum gravity (QG), which may source the observed cosmic acceleration through interactions among the quantum constituents of the spacetime~\cite{Oriti:2021rvm}. In this picture, the breakdown of the smooth fabric of space at the Planck scale influences the dynamics of the large-scale structures and, consequently, the behavior of the systems within the Universe. Non-trivial effects may manifest in the fundamental properties of position and momentum measurements and therefore their uncertainty relations.

To account for this QG phenomenology, several studies have proposed modifying the Heisenberg uncertainty relation of the quantum theory into the so-called \emph{generalized uncertainty principle} (GUP)~\cite{Amati:1987wq,Gross:1987kza,Amati:1988tn,Konishi:1989wk,Maggiore:1993kv,Capozziello:1999wx,Kempf:1994su,Scardigli:1999jh,Adler:2001vs, Hossenfelder:2012jw,Nouicer:2007jg,Park:2007az,Ong:2018zqn,Buoninfante:2019fwr,Buoninfante:2020cqz,Chemisana:2023fuk,Das:2008kaa,Bosso:2017ndq,Luciano:2019mrz,Luciano:2021ndh,Jizba:2023ygi,Buoninfante:2020guu,Bosso:2021koi,Iorio:2022ave,Pikovski:2011zk,Addazi:2021xuf,AlvesBatista:2023wqm}. The most common expression of the GUP includes a quadratic correction in the momentum uncertainty of the form~\cite{Kempf:1994su}
\be
\label{GUP}
\Delta x\Delta p\ge \frac{1}{2}\left(1+\beta l_p^2\Delta p^2\right),
\ee
where $\Delta x$, $\Delta p$ are the usual position and momentum uncertainties, respectively, while $\beta$ is the dimensionless GUP parameter. 
The inequality~\eqref{GUP} admits a minimum uncertainty $\Delta x\sim\sqrt{\beta}\,l_p$ for $\beta>0$, which is the case we shall consider in the present analysis. However, models with $\beta<0$ have also been studied~\cite{Jizba:2009qf,Ong:2018zqn,Buoninfante:2019fwr} (see~\cite{Bosso:2023aht} for a recent review).  
The standard quantum mechanics is recovered for $\beta=0$ and/or $l_p^2\Delta p^2\ll1$, i.e., at energy scales far enough from the Planck threshold. 

%Theoretical implications of the GUP have been  investigated in the natural contexts of black hole physics~\cite{Adler:2001vs, Hossenfelder:2012jw,Nouicer:2007jg,Park:2007az,Jizba:2009qf,Carr:2015nqa,Ong:2018zqn,Casadio:2013aua,Buoninfante:2019fwr,Buoninfante:2020cqz,Alonso-Serrano:2018ycq,Chemisana:2023fuk}, quantum theory~\cite{Das:2008kaa,Frassino:2011aa,Bosso:2017ndq,Luciano:2019mrz,Luciano:2021ndh,Jizba:2023ygi,Buoninfante:2020guu,Bosso:2022vlz,Bosso:2021koi}, condensed matter~\cite{Iorio:2022ave} and optics~\cite{Pikovski:2011zk}. Additionally, a branch of research within QG phenomenology aims to indirectly constrain the GUP by studying its low-energy effects (see~\cite{Bosso:2023aht} for a review). 

Due to the semiclassical nature of the GUP, observational implications are naturally expected in astrophysics and cosmology in the quasi-classical quantum domain. 
However, to the best of our knowledge, there are very few studies that deal with the application of Eq.~\eqref{GUP} to these frameworks. For instance, the recent analysis in~\cite{Jizba:2022icu} has pointed out that the coherent states of the GUP coincide with the Tsallis' probability amplitude, opening new perspectives in conformal gravity. Further investigation in the cosmological realm appears in~\cite{Zhu:2008cg,Giardino:2020myz,Luciano:2021vkl,Luciano:2022ely}. 

Building upon these premises, in this work
consider a semiclassical cosmological model arising from GUP-modified Friedmann equations. In this extended scenario, we find that the GUP behaves as an effective quintessence-like DE that adds to the cosmological constant, influencing the early Universe's dynamics while recovering the $\Lambda$CDM in the present era. From this perspective, our result aligns with previous achievements in the literature~\cite{Maggiore:2010wr,Kim:2008hz,Lake:2017uzd,Kouwn:2018rmp}, which argue that the GUP correction would dilute too quickly to account for the observed DE budget on its own.
In other terms, while the GUP cosmology may effectively model QG effects in the primordial Universe, it still requires a non-vanishing cosmological constant to explain late-time phenomenology.

We use the above findings to disclose the impact of Eq.~\eqref{GUP} on some characteristic cosmological parameters, such as the Equation of State (EoS) parameter of DE and the Hubble rate.
We then explore the growth of matter perturbations in the Top-Hat Spherical Collapse (SC) approach~\cite{Abramo:2007iu}. This model involves considering a uniform, spherically symmetric perturbation within an expanding background, which is described using the same Friedmann equations as those in the underlying theory of gravity~\cite{Planelles:2014zaa}. 
Finally, we analyze implications for Primordial Gravitational Waves (PGWs). %It is known that the propagation of PGWs carries unique information on the expansion history of the Universe through its evolution in the post-inflationary phase. Hence, such GWs represent a useful tool to probe cosmological events prior to BBN~\cite{Gouttenoire:2019kij,Auclair:2019wcv,Garcia-Bellido:2007fiu,Hajkarim:2019csy,Hajkarim:2019nbx,Giudice:2016zpa,Bernal:2020ywq}. 
We identify the $\beta$-parameter space that enhances the PGW spectrum, offering a test to detect/constrain quantum-induced deviations from General Relativity (GR) by upcoming GW observatories.
In that regard, the current effort differs from~\cite{Das:2022hjp}, which explores GUP effects on PGWs through the connection with modified gravity. 
On the other hand, we provide a complementary perspective to that proposed in~\cite{Marin:2013pga,Bhattacharyya:2023xvv}, where QG corrections are constrained by considering resonant bar detector of GWs.

The structure of the work is as outlined below: following~\cite{Kouwn:2018rmp}, in Sec.~\ref{TB} we investigate the influence of the GUP on the Friedmann equations for the
FRW Universe. Implications for the evolution of primordial matter fluctuations are studied in Sec.~\ref{Gro}, while Sec.~\ref{GW} is devoted to PGW analysis. Conclusions are summarized in Sec.~\ref{Co}.

\section{Theoretical Framework}
\label{TB}
Let us start by reviewing the computation of the modified Friedmann equations in the GUP cosmology. Toward this end, our strategy is to first derive the GUP-corrected Bekenstein-Hawking entropy for the black hole horizon. The resulting expression is translated into a deformed entropy-area law, which is then extended to the apparent horizon of the Universe~\cite{Jacobson:1995ab}. %This is justified by the fact that any causal horizon is inevitably associated with entropy, as originally pointed out in~\cite{Jacobson:1995ab}. 
Using the gravity-thermodynamics conjecture, we finally obtain Friedmann equations within Jacobson's approach. \textcolor{black}{
In this context, it is important to highlight that the general formula for the modified cosmological equations, derived using a quantum corrected entropy-area relation, was presented in~\cite{Cai:2008ys}.}

From Eq.~\eqref{GUP}, it is straightforward to verify that
\be
\label{GUP2}
\Delta p\ge \frac{1}{2\Delta x} \left(1+\frac{\beta l_p^2}{4\Delta x^2}\right),
\ee
where we have expanded up to $\mathcal{O}\left(\beta l_p^2/\Delta x^2\right)$. Following~\cite{Kouwn:2018rmp}, we apply this inequality to a test particle absorbed/emitted by a black hole. Denoting by $E$ and $R$ the energy and size of the particle, respectively, the surface area of the black hole will change by $\Delta A\gtrsim 8\pi l_p^2 E R$~\cite{Christodoulou:1970wf}, up to a calibration factor that will be fixed below. 
Since the size of the particle cannot be smaller than $\Delta x$~\cite{Amelino-Camelia:2005zpp}, the minimal change of the
black hole area satisfies $\Delta A_{min}\gtrsim8\pi l_p^2 E \Delta x$. 
Translating the inequality~\eqref{GUP2} into a lower bound on $E$~\cite{Medved:2004yu}, we then obtain
\be
\label{GUP3}
\Delta A_{min}\gtrsim4\pi l_p^2\left(1+\frac{\beta l_p^2}{4\Delta x^2}\right).
\ee

We further assume that the particle is located around the black hole with a position uncertainty $\Delta x\simeq2r_s$,  where $r_s$ is the Schwarzschild radius and $A=4\pi r_s^2$ the horizon area of the black hole. We can then write $A=\pi\Delta x^2$, which gives 
\be
\label{GUP4}
\Delta A_{min}\ge4\pi\lambda l_p^2\left(1+\frac{\beta\pi l_p^2}{4A}\right),
\ee
where we have included the calibration factor $\lambda=\Delta S_{min}/\pi$, with $\Delta S_{min}=\log 2$ being the minimal entropy increase associated to the absorption/emission of one bit of information~\cite{Kouwn:2018rmp,Awad:2014bta}.

The minimal change of
entropy associated to the minimal change in the horizon area reads
\be
\label{GUP5}
\frac{dS}{dA}=\frac{\Delta S_{min}}{\Delta A_{min}}=\frac{1}{4l_p^2}\left(1+\frac{\beta\pi l_p^2}{4A}\right)^{-1},
\ee
which can be integrated to give~\cite{Kouwn:2018rmp}
\be
\label{GUP6}
S_{\beta}(A)\simeq\frac{A}{4l_p^2}\left[1-\frac{\beta\pi l_p^2}{4A}\log\left(\frac{A}{4l_p^2}\right)\right], 
\ee
up to an additional constant depending on the normalization adopted. Nevertheless, this factor can be safely neglected, as only derivatives of $S_\beta$ will appear in our next analysis.   

The relation~\eqref{GUP6} provides the generalized Bekenstein-Hawking entropy in the presence of the GUP. For $\beta=0$, we recover the standard holographic scaling. It is noteworthy that logarithmic-like deformations of the entropy-area law have been obtained in
a number of different approaches to QG, including string theory, AdS/CFT correspondence and loop quantum gravity. Therefore, we expect our next considerations and results to have general validity within the QG framework.

\subsection{GUP-modified Friedmann equations}
Due to the geometrical - and thus universal - nature of the law~\eqref{GUP6}, it does not only apply to black holes but can also be extended to the apparent horizon of the Universe to develop a GUP-modified cosmology~\cite{wang2001relating,Cai:2005ra,PhysRevLett.96.121301,Akbar:2006er,Akbar:2006mq,Frolov:2002va,Sheykhi:2007zp}. Without loss of generality, we focus on a flat ($k=0$) Friedmann-Robertson-Walker  (FRW) Universe of background metric
\be
ds^2 \ = \ h_{bc} \ \! dx^b dx^c \ + \ \tilde r^2\left(d\theta^2 \ + \ \sin^2\theta\, d\phi^2\right) ,
\label{FRW}
\ee
where $\tilde r=a(t)\hspace{0.2mm}r$, $x^0=t$, $x^1=r$, $h_{bc}=\mathrm{diag}\left(-1,a^2\right)$ and $a(t)$ is the time-dependent scale factor.  

For the FRW Universe,  the apparent horizon is a non-expanding, marginally trapped surface that always exists, unlike the event or particle horizon. Furthermore, it can be characterized similarly to the event horizon of black holes, thus providing the most suitable cosmological horizon for thermodynamic analysis~\cite{Cai:2005ra,Faraoni:2011hf}. The defining relation is $h^{ab}\partial \tilde r_a\partial\tilde r_b=0$, which gives for the apparent horizon radius $\tilde r_A^2=1/H^2$, where $H=\dot a/a$ is the Hubble rate and the overdot denotes time derivative. 

We assume the energy content of the Universe is a perfect fluid. Denoting its four-velocity by $u^\mu$, the energy-momentum tensor takes the form $T_{\mu\nu}=\left(\rho+p\right)u_\mu u_\nu+pg_{\mu\nu}$, where $\rho$ and $p$ are
the energy density and pressure of the fluid. In turn, the energy conservation law $\nabla_\mu T^{\mu\nu}=0$ implies the continuity equation $\dot\rho=-3H\left(\rho+p\right)$.

According to the gravity-thermodynamics conjecture, the whole Universe behaves as a thermodynamic system bounded by the
apparent horizon $\tilde r_A$. In turn,  
the cosmological equations can be derived from the first law of thermodynamics applied to the bounding surface~\cite{Jacobson:1995ab,Cai:2005ra,PhysRevLett.96.121301,Akbar:2006er,Akbar:2006mq,Frolov:2002va,Sheykhi:2007zp}. Following~\cite{Akbar:2006er,Awad:2014bta,Kouwn:2018rmp}, we obtain 
\begin{eqnarray}
\label{F1}
\frac{8\pi G}{3}\rho&=&-16\pi G\int \frac{\partial_{A} S(A)}{A^2}dA\,,\\[0.5mm]
\label{F2}
-\pi\left(\rho+p\right)&=&\partial_{A}S(A)\,\dot H\,,
\end{eqnarray}
where $G=l_p^2$ is the gravitational constant. 

The dynamics of the Universe depends on the explicit form of the horizon entropy $S(A)$.
Within the specific GUP framework, using Eq.~\eqref{GUP6}, we can write~\cite{Kouwn:2018rmp}
\begin{eqnarray}
\label{HuPa}
H^2&=&\frac{1}{3M_p^2}\left(\rho+\rho_{DE}\right),\\[0.5mm]
\dot H&=&-\frac{1}{2M_p^2}\left(\rho+p+\rho_{DE}+p_{DE}\right),
\end{eqnarray}
where $M_p^2=1/(8\pi G)$ is the reduced Planck mass and 
\begin{eqnarray}
\label{F3}
    \rho_{DE}&=&M_p^2\Lambda+\tilde \beta H^4\,,\\[1mm]
    \label{F4}
    p_{DE}&=&-\left[ M_p^2\Lambda+\tilde\beta H^2\left(\frac{4}{3}\dot H+H^2\right)\right].
\end{eqnarray}
Here, $\tilde\beta\equiv3\beta/(256\pi)$, while the integration constant $\Lambda$ plays the role of the cosmological constant. The standard Friedmann equations are obtained for $\beta=0$, as expected. 

Interestingly, the GUP introduces a dynamical DE component in addition to the cosmological constant. We stress that the scenario without an explicit $\Lambda$ would be physically inconsistent with late-time phenomenology. Indeed, since the additional term scales as $\tilde\beta H^4$, it would dilute too quickly to account for the current cosmic acceleration on its own. This conclusion aligns with the findings of~\cite{Maggiore:2010wr}, even for a slower scaling $\rho_{DE}\sim H^2$. A different perspective is offered by~\cite{Paliathanasis:2015cza}, which assumes a gravitational langrangian with a quintessence scalar field modified by the GUP.

\subsection{Cosmological implications}
Let us elaborate on the implications of the GUP cosmology. From Eqs.~\eqref{F3}-\eqref{F4}, the effective EoS of DE reads
\begin{equation}
\label{Eos}
    w_{DE}=\frac{p_{DE}}{\rho_{DE}}=-1-\frac{4\tilde\beta H^2\dot H}{3\left(M_p^2\Lambda+\tilde\beta H^4\right)}.
\end{equation}

We assume the Universe is filled with a perfect fluid consisting of  pressureless matter ($m$) and radiation ($r$). \textcolor{black}{In turn, the Hubble parameter in Eq.~\eqref{HuPa} can be written as
\begin{equation}
\label{15}
H^2(z)=\frac{8\pi G}{3}\left[\rho_{m,0}\left(1+z\right)^3+\rho_{r,0}\left(1+z\right)^4+\rho_{DE}(z)\right],
\end{equation}
%Defining for each component the dimensionless densities $\Omega_i=8\pi G\hspace{0.2mm} \rho_i/3H^2\, (i=m,r,DE)$,
where we have used the definition of the reduced Planck mass $M_p$ introduced above Eq.~\eqref{F3}, and expressed the energy density $\rho$ of the fluid as the sum of its matter ($\rho_m$) and radiation ($\rho_{r}$) contributions, respectively.
Furthermore, we have used the scale-dependence $\rho_m (a)=\rho_{m0}/a^3$,  $\rho_r (a)=\rho_{r0}/a^4$ and introduced the redshfit-dependence $1+z=1/a$. The subscript ``0'' denotes the present value of the corresponding quantity (we have implicitly assumed $a_0=1$). }

\textcolor{black}
{Equation~\eqref{15} can be reformulated in terms of the fractional energy densities $\Omega_i=8\pi G\hspace{0.2mm} \rho_i/(3H^2)\, (i=m,r,DE)$ by multiplying and dividing the first two terms by $H_0^2$ and the last term by $H^2$. By doing so, we obtain
\begin{eqnarray}
\nonumber
&&\hspace{-5mm}H^2(z)=H_0^2\left[\Omega_{m,0}\left(1+z\right)^3+\Omega_{r,0}\left(1+z\right)^4\right]+\Omega_{DE}(z) H^2(z)\\[2mm]
&&\hspace{-3mm}\Longrightarrow \,H(z)=H_0\sqrt{\frac{\Omega_{m0}\left(1+z\right)^3+\Omega_{r0}\left(1+z\right)^4}{1-\Omega_{DE}(z)}}\,.
\label{Hrate}
\end{eqnarray}
}

Upon substitution of Eq.~\eqref{F3} into the definition of $\Omega_{DE}$, we acquire 
\begin{widetext}
\begin{eqnarray}
\label{OmegaDE}
\nonumber
\hspace{-8mm}\Omega_{DE}(z)&=&\frac{1}{2\left\{\Lambda+3H_0^2\left[\Omega_{m0}\left(1+z\right)^3+\Omega_{r0}\left(1+z\right)^4\right]\right\}}\Bigg\{{2\Lambda+3H_0^2\left[\Omega_{m0}\left(1+z\right)^3+\Omega_{r0}\left(1+z\right)^4\right]}\\[1mm]
&&-H_0^2\left(1+z\right)^2\left[\Omega_{m0}\left(1+z\right)^3+\Omega_{r0}\left(1+z\right)^4\right]\sqrt{\frac{9-32\pi \tilde\beta G\left\{\Lambda+3H_0^2\left[\Omega_{m0}\left(1+z\right)^3+\Omega_{r0}\left(1+z\right)^4\right]\right\}}{\left(1+z\right)^4}}\Bigg\},
\end{eqnarray}
\end{widetext}
where the constant $\Lambda$ is fixed by the initial condition $\Omega_{m0}+\Omega_{r0}+\Omega_{DE,0}=1$, which implies 
\begin{equation}
\label{Lambda}
    \Lambda=H_0^2\left[3\left(1-\Omega_{m0}-\Omega_{r0}\right)-8\pi\tilde\beta G H_0^2
    \right].
\end{equation}

\begin{figure}[t]
\centering
\includegraphics[width=8cm]{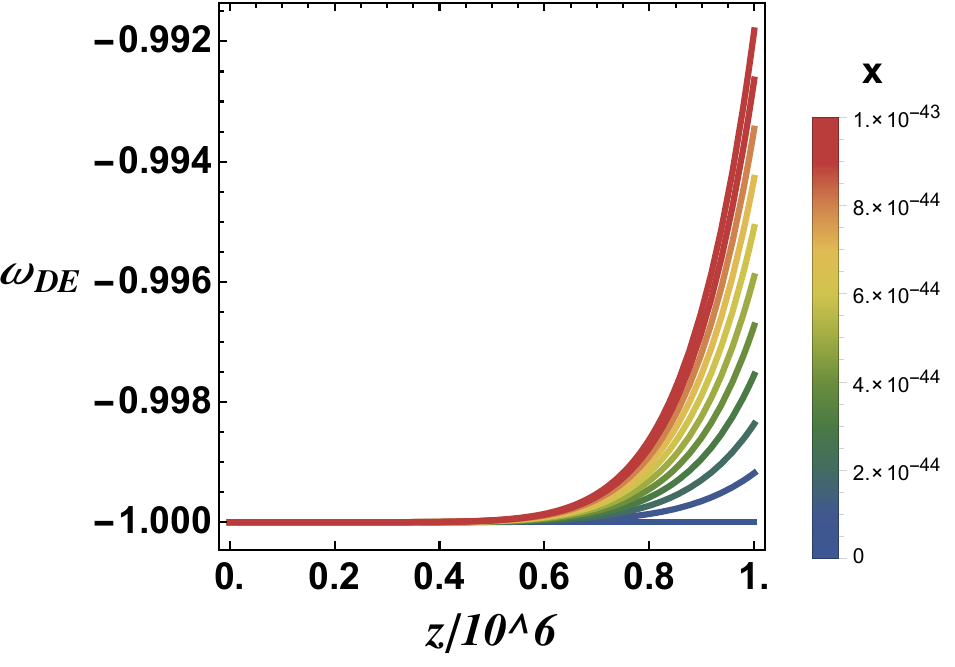}
\caption{Plot of $\omega_{DE}$ versus $z$, for various values of $x$ and fixed $\Omega_{m0}\simeq0.3, \Omega_{r0}\simeq5\times10^{-5}$.}
\label{Fig1}
\end{figure}

For our next purposes, it proves convenient to introduce the dimensionless parameter $x=\tilde\beta G H_0^2$. Since $G\simeq 10^{-38}\,\mathrm{GeV}^{-2}$, $H_0\simeq 10^{-42}\,\mathrm{GeV}$, it is reasonable to expand around $x\ll1$ as long as $\tilde\beta\ll 10^{122}$. This is certainly true in string theory, where the GUP parameter is assumed to be of order unity, but also in cosmological and astrophysical contexts for consistency with the latest experimental data~\cite{Giardino:2020myz,Bosso:2023aht}.
The energy density in Eq.~\eqref{OmegaDE} then becomes
\begin{eqnarray}
\nonumber
    \Omega_{DE}(z)&\approx& \frac{\Lambda}{\Lambda+3H_0^2\left[\Omega_{m0}\left(1+z\right)^3+\Omega_{r0}\left(1+z\right)^4\right]}\\[1mm]
    &&+\frac{8\pi x}{3}\left[\Omega_{m0}\left(1+z\right)^3+\Omega_{r0}\left(1+z\right)^4\right],
    \label{OmegaDEbis}
\end{eqnarray}
where we have expanded up to $\mathcal{O}(x)$ to be consistent with the linear GUP~\eqref{GUP}.  Accordingly, the EoS~\eqref{Eos} reads
\begin{eqnarray}
\nonumber
    \omega_{DE}&\approx&-1+\frac{16\pi x}{9}\frac{\left[3\Omega_{m0}\left(1+z\right)^3+4\Omega_{r0}\left(1+z\right)^4\right]}{\left(1-\Omega_{m0}-\Omega_{r0}\right)}\\[1mm]
    &&\hspace{-12mm}\times
\left\{1-\Omega_{m0}\left[1-\left(1+z\right)^3\right]-\Omega_{r0}\left[1-\left(1+z\right)^4\right]\right\},
\end{eqnarray}
where we have explicitly substituted $\Lambda$ in Eq.~\eqref{Lambda}.

The evolution of $\omega_{DE}$ versus $z$ is displayed in Fig.~\ref{Fig1} for various values of $x$. As expected, the GUP correction to DE appears more prominent at higher redshift, where the quantum nature of gravity should significantly influence the dynamics of the Universe. In this regime, the GUP manifests as an effective quintessence-like component, i.e., $-1<\omega_{DE}<-1/3$. However, as $z$ decreases, this extra term rapidly dilutes, recovering the cosmological constant behavior $(\omega_{DE}=-1)$ in the current epoch.

Substituting Eq.~\eqref{OmegaDEbis} into~\eqref{Hrate}, we can also derive the modified expression of the Hubble rate
\begin{widetext}
\begin{equation}
\label{H}
\frac{H(z)}{H_0}\approx \sqrt{1-\Omega_{m0}\left[1-\left(1+z\right)^3\right]-\Omega_{r0}\left[1-\left(1+z\right)^4\right]}
\,-\,\frac{4\pi x}{3}\,\frac{1-\left\{1-\Omega_{m0}\left[1-\left(1+z\right)^3\right]-\Omega_{r0}\left[1-\left(1+z\right)^4\right]\right\}^2}{\sqrt{1-\Omega_{m0}\left[1-\left(1+z\right)^3\right]-\Omega_{r0}\left[1-\left(1+z\right)^4\right]}}.
\end{equation}
\end{widetext}

\begin{figure}[t]
\centering
\includegraphics[width=9.1cm]{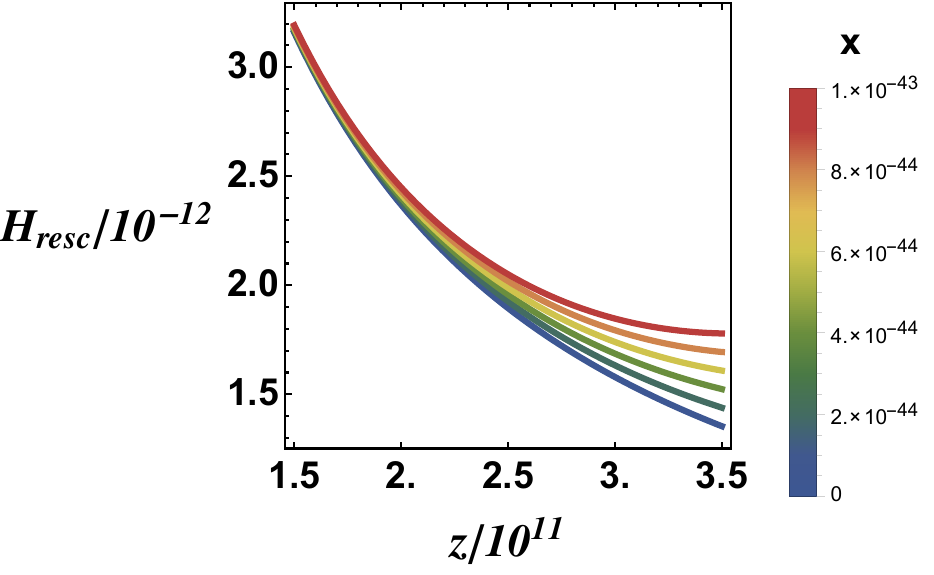}
\caption{Plot of $H_{resc}$ versus $z$, for various values of $x$ and fixed $\Omega_{m0}\simeq0.3, \Omega_{r0}\simeq5\times10^{-5}$. We have restored physical units and expressed the Hubble rate in $\mathrm{Km/s/Mpc}$.}
\label{Fig2}
\end{figure}

Figure~\ref{Fig2} displays the behavior of $H_{resc}(z)\equiv H/(1+z)^3$ for various values of $x$. We can see that the presence of the GUP speeds up the expansion of the Universe at higher redshift, in such a way that, the larger $x$, the higher the departure from the standard $\Lambda$CDM evolution (blue curve). Further discussion on the physical interpretation of this result is provided in the next section, in connection with the study of the gravitational evolution of matter perturbations and structure formation. 
On the other hand, $H(z)\rightarrow H_0$ for $z\rightarrow0$, as one can straightforwardly check from Eq.~\eqref{H}. 

\section{Growth of Matter perturbations}
\label{Gro}
A major challenge in modern cosmology involves understanding the origin and development of matter perturbations in the early Universe. It is widely accepted that the large-scale structures we observe today originated from the gravitational evolution of small primordial fluctuations in the matter density. Such fluctuations grew over time until they became strong enough to decouple from the background expansion and collapse into gravitationally bound systems~\cite{Mukhanov:1990me}. A useful framework in this context is the Top-Hat Spherical Collapse (SC) model~\cite{Abramo:2007iu}, which analyzes uniform, spherically symmetric perturbations in an expanding background and describes their evolution within a spherical region using the Friedmann equations of gravity.

To explore the impact of the GUP on the growth mechanism of perturbations, let us rewrite the second modified Friedmann equation as
\begin{equation}
\label{pertmodeq}
    \frac{\ddot a }{a}\approx\frac{1}{3}\left(\Lambda-4\pi G \rho_m\right)\,+\,\frac{8\pi x}{27H_0^2}\left(\Lambda-16\pi G\rho_m\right)\left(\Lambda+8\pi G\rho_m\right),
\end{equation}
where we have used $\dfrac{\ddot a }{a}=\dot H+H^2$.

We further consider a spherical region of radius $a_p$ and homogeneous density $\rho_c$. Denoting by $\delta\rho_m$ the density fluctuation inside the region, we can write 
\begin{equation}
\rho_c=\rho_m+\delta\rho_m\,.
\end{equation}
In turn, the conservation equation reads $\dot\rho_c+3H_p\rho_c=0$, 
where $H_p\equiv\dot a_p/a_p$ is the Hubble rate inside the spherical region~\cite{Abramo:2007iu}. 

In terms of the perturbation $\delta\rho_m$, the density contrast is defined as
\begin{equation}
\label{cont}
    \delta_m\equiv\frac{\rho_c-\rho_m}{\rho_m}=\frac{\delta\rho_m}{\rho_m}\,.
\end{equation}
The evolution equation for the matter perturbations can be obtained by deriving Eq.~\eqref{cont} with respect to $t$ and using the continuity equation. In so doing, we are led to $\dot\delta_m=3\left(1+\delta\right)\left(H-H_p\right)$, which can be further differentiated to give
\begin{equation}
\label{evol}
    \ddot\delta_m=3\left(1+\delta_m\right)(\dot H-\dot H_p)+\frac{\dot\delta_m^2}{1+\delta_m}\,.
\end{equation}

Since the early stages of the gravitational
collapse are well described by tiny fluctuations for all but the very smallest sales, it is reasonable to assume $\delta\rho_m\ll\rho_m\Longrightarrow\delta_m\ll1$~\cite{Abramo:2007iu}. In the linear regime, noting that
\begin{equation}
  \hspace{-1mm}  \dot H-\dot H_p\approx H_p^2-H^2\,+\,\frac{4\pi G}{3}\rho_m\left[1+\frac{16\pi x}{9H_0^2}\left(\Lambda+32\pi G\rho_m\right)\right]\delta_m\,,
\end{equation}
the dynamics~\eqref{evol} becomes
\begin{equation}
\label{evol2}
    \ddot\delta_m+2H\dot\delta_m-4\pi G\rho_m\left[1+\frac{16\pi x}{9H_0^2}\left(\Lambda+32\pi G\rho_m\right)\right]\delta_m=0\,.
\end{equation}
This relation rules the growth of matter perturbations under the influence of the GUP. 

It is useful to parameterize the evolution of $\delta_m$ in terms of the scale factor (or, equivalently, the redshift). Using the prime notation for the derivative with respect to $a$, we have
\begin{eqnarray}
    \dot\delta_m&=&Ha\hspace{0.4mm}\delta_m'\,,\\[1mm]
    \ddot\delta_m&=&H^2a^2\delta_m''+Ha\left(H'a+H\right)\delta_m'\,.
\end{eqnarray}
Upon substitution into Eq.~\eqref{evol2}, we get
\begin{equation}
\label{evoln}
    \delta_m''+\frac{\delta_m'}{a}\left(\frac{3}{2}-4\pi x\,\frac{\Omega_{m0}}{a^3}\right)-\frac{\delta_m}{a^2}\left(\frac{3}{2}+28\pi x\,\frac{\Omega_{m0}}{a^3}\right)=0\,,
\end{equation}
where we have omitted negligible cosmological constant effects. Notice that 
this is not at odds with our former assumption, since we are here considering the primordial evolution of the Universe far from the current epoch. The differential equation~\eqref{evoln} reproduces the standard dynamics when $\beta=0$~\cite{Abramo:2007iu}. 

\begin{figure}[t]
\centering
\includegraphics[width=8.5cm]{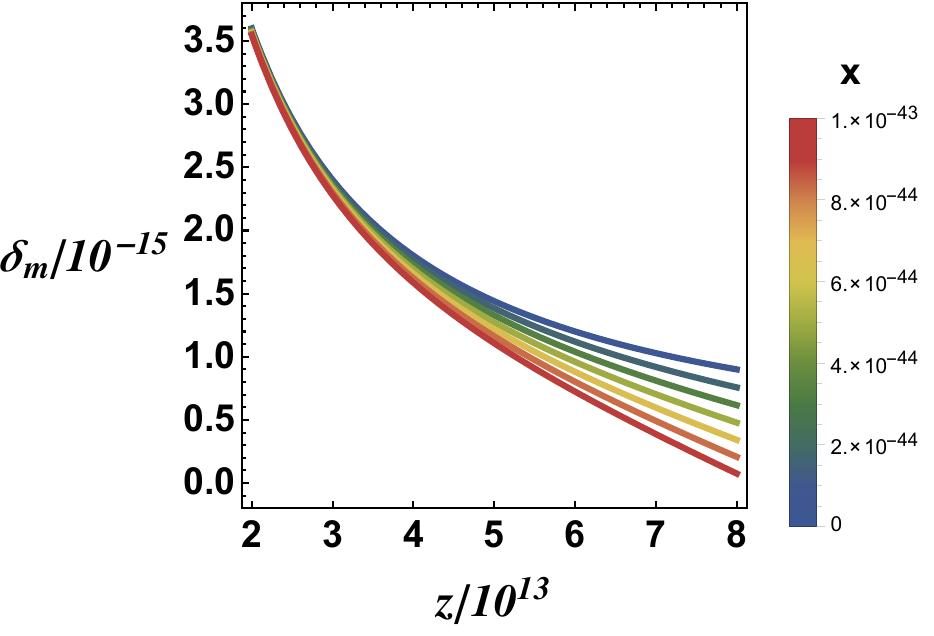}
\caption{Plot of the growth of matter perturbations (density contrast) versus $z$, for various values of $x$ and fixed $\Omega_{m0}\simeq0.3$. We set the same initial conditions as in~\cite{Abramo:2007iu}.}
\label{Fig4}
\end{figure}

We have solved Eq.~\eqref{evoln} numerically.
The density profile is plotted in Fig.~\ref{Fig4} for various values of $x$. Compared to the standard cosmology (blue curve), the GUP smooths out small-scale fluctuations in the early Universe. 
In turn,  the growth of perturbations is  suppressed, resulting in 
a slower increase of the density contrast and a delayed formation of large-scale structures. As a consequence, the gravitational pull that would normally slow down the expansion turns out to be weaker, allowing the Universe to expand faster in the initial stages.
This result is in line with the profile of the Hubble rate in Fig.~\ref{Fig2}.
We mention that a similar outcome has been obtained in~\cite{Hamber:2014dea}, assuming a running gravity coupling in QG.

Despite the non-trivial finding, further work may be required, particularly to extend the above analysis beyond the linear regime of small density contrast.

\section{Primordial Gravitational Waves}
\label{GW}

The groundbreaking detection of gravitational waves (GWs)~\cite{LIGOScientific:2014pky,VIRGO:2014yos}  has opened new opportunities for exploring the Universe. GW signals are conventionally classified into three primary sources based on their generation mechanisms: astrophysical, cosmological, and inflationary sources. Astrophysical GWs are significantly influenced by the mass of the emitting object. Particularly, a minimum mass $M\sim M_\odot$ would correspond to a maximum frequency of around $f\simeq 10\,\mathrm{kHz}$.
In contrast, various early Universe cosmological events, such as first-order phase transitions at approximately $10^{-2}\, \mathrm{GeV}$ or just below the Grand Unified Theory scale, can generate GWs with frequencies around $f\simeq10^{-5}\, \mathrm{Hz}$ or lower than the $\mathrm{GHz}$ range, respectively. Additionally, primordial GWs produced during inflation cover a wide range of frequencies from $10^{-18}\,\mathrm{Hz}$ to $1\, \mathrm{GHz}$~\cite{Ito:2022rxn,Jizba:2024klq}. 

While current experiments are mostly focused on astrophysical signals, the future detection of high-frequency GWs may offer a promising yet challenging avenue for testing theories beyond the $\Lambda$CDM. In this perspective,  we compute the spectrum of primordial gravitational waves (PGWs) in the framework of GUP-cosmology. Concretely, we consider signals in the frequency range below $10^3\,\mathrm{Hz}$, which is expected to be fully probed by current and upcoming GW observatories (see Fig.~\ref{Fig5}). A similar study has recently appeared in~\cite{Odintsov:2024sbo} in the context of a super-generalized (but still classical) entropy.

\subsection{Standard cosmology}
To set up the notation, let us briefly review the computation of the relic density of PGW in GR. In the linearized approximation, GWs are modeled as metric perturbations $h_{\mu\nu}$ on a curved spacetime. We consider waves propagating on an isotropic, uniform and flat background, so that $h_{00} =  h_{0i} = 0$. 
Using the transverse traceless
($TT$) gauge, i.e., $\partial^i h_{ij}=0$ and $h^i_i=0$, the dynamics of the tensor perturbations in the first-order perturbation theory obeys~\cite{Watanabe:2006qe}
\be
\label{hdyn}
\ddot h_{ij}\ + \ 3H\dot h_{ij} \ - \ \frac{\nabla^2}{a^2}h_{ij}\ = \ 16\pi G\hspace{0.3mm} \Pi_{ij}^{TT}\,,
\ee
where 
\be
\Pi_{ij} = \frac{T_{ij}-p\hspace{0.3mm} g_{ij}}{a^2}\,,
\label{TPE}
\ee
is the $TT$ anisotropic part of the stress tensor $T_{ij}$, $g_{ij}$ the metric tensor
and $p$ the background pressure (in our convention, latin
indices run over the spatial coordinates).

To study the dynamics~\eqref{hdyn}, we follow~\cite{Bernal:2020ywq} and work in the Fourier space, where
\be
h_{ij}(t,\vec{x}) = \sum_{\lambda}\int\frac{d^3k}{\left(2\pi\right)^3}\hspace{0.2mm}h^\lambda(t,\vec{k})\hspace{0.2mm}\epsilon^\lambda_{ij}(\vec{k})\hspace{0.2mm}e^{i\vec{k}\cdot\vec{x}}\,.
\ee
Here,  $\epsilon^\lambda$ is the spin-2
polarization tensor, satisfying 
$\sum_{ij}\epsilon^\lambda_{ij}\epsilon^{\lambda'*}_{ij}=2\delta^{\lambda\lambda'}$. The index $\lambda$ runs over the two independent wave-polarizations $\lambda=+,\times$, while the  over-arrow is the usual notation for three-vectors. 

It proves convenient to factorize the time-dependence of the tensor perturbation $h^\lambda(t,\vec{k}) = h_{\mathrm{prim}}^\lambda(\vec{k})\,X(t,k)$,
where $k\equiv|\vec{k}|$,  $h_{\mathrm{prim}}^\lambda(\vec{k})$ 
is the amplitude of the primordial tensor perturbation and $X(t,k)$ the transfer function containing its time-dependence.  

With the above tools at hand, the tensor power spectrum of GWs is~\cite{Bernal:2020ywq}
\be
\label{TPS}
\mathcal{P}_T(k) \ = \ \frac{k^3}{\pi^2}\sum_\lambda\Big|h^\lambda_{\mathrm{prim}}(\vec k)\Big|^2 \ = \ \frac{2}{\pi^2}\hspace{0.3mm}G\hspace{0.3mm} H^2\Big|_{k=aH}\,.
\ee
Moreover, Eq.~\eqref{hdyn} takes the damped harmonic oscillator-like form
\be
\partial^2_\tau X \ + \ 2\hspace{0.2mm}\frac{   \partial_\tau a}{a}\,\partial_\tau X \ + \ k^2X \ = \ 0\,,
\ee
where we have introduced the conformal time $d\tau=dt/a$.

The relic density from first-order tensor perturbation is defined by~\cite{Watanabe:2006qe}
\begin{eqnarray}
\nonumber
\Omega_{\mathrm{GW}}
(\tau,k)&=&\frac{[\partial_{\tau}X(\tau,k)]^2}{12a^2(\tau)H^2(\tau)}\,\mathcal{P}_T(k)\\[1mm]
&\simeq&\left[\frac{a_{\mathrm{hc}}}{a(\tau)}\right]^4\left[\frac{H_{\mathrm{hc}}}{H(\tau)}\right]^2\frac{\mathcal{P}_T(k)}{24}\,,
\label{Ttps}
\end{eqnarray}
where $H_{hc}$ and $a_{hc}$ are the Hubble rate and scale factor at the horizon crossing, respectively. In the second step, we have averaged over oscillation periods, i.e., 
\be
X'(\tau,k)\ \simeq \ k\hspace{0.2mm} X(\tau,k)\ \simeq \  \frac{k\hspace{0.3mm} a_{\mathrm{hc}}}{\sqrt{2}a(\tau)}\ \simeq \
\frac{a^2_{\mathrm{hc}}\hspace{0.3mm}H_{\mathrm{hc}}}{\sqrt{2}a(\tau)}\,,
\ee
where 
\begin{equation}
\label{freq}
k=2\pi f = a_{hc}H(a_{hc})
\end{equation}
at the horizon crossing~\cite{Bernal:2020ywq}.

Using Eq.~\eqref{Ttps}, the current PGW relic density reads
\begin{equation}
\Omega_{\mathrm{GW}}(\tau_0,k)h^2 \simeq\left[\frac{g_*(T_{\mathrm{hc}})}{2}\right]\left[\frac{g_{*s}(T_0)}{g_{*s}(T_{\mathrm{hc}})}\right]^{4/3}\frac{\mathcal{P}_T(k)\Omega_{r}(T_0)h^2}{24}\,,
\label{Spt}
\end{equation}
where $h$ is the reduced Hubble constant and $\Omega_{r}(T_0)$ the dimensionless radiation density at present temperature $T_0$. The effective numbers $g_{*}$, $g_{*s}$ of relativistic degrees of freedom that contribute to the radiation energy density $\rho_r$ and entropy density $s_r$ are defined through $\rho_r=\pi^2 g_*(T)T^4/30$ and $s_r=2\pi^2g_{*s}(T)T^3/45$.

\begin{figure}[t]
\centering
\includegraphics[width=8.5cm]{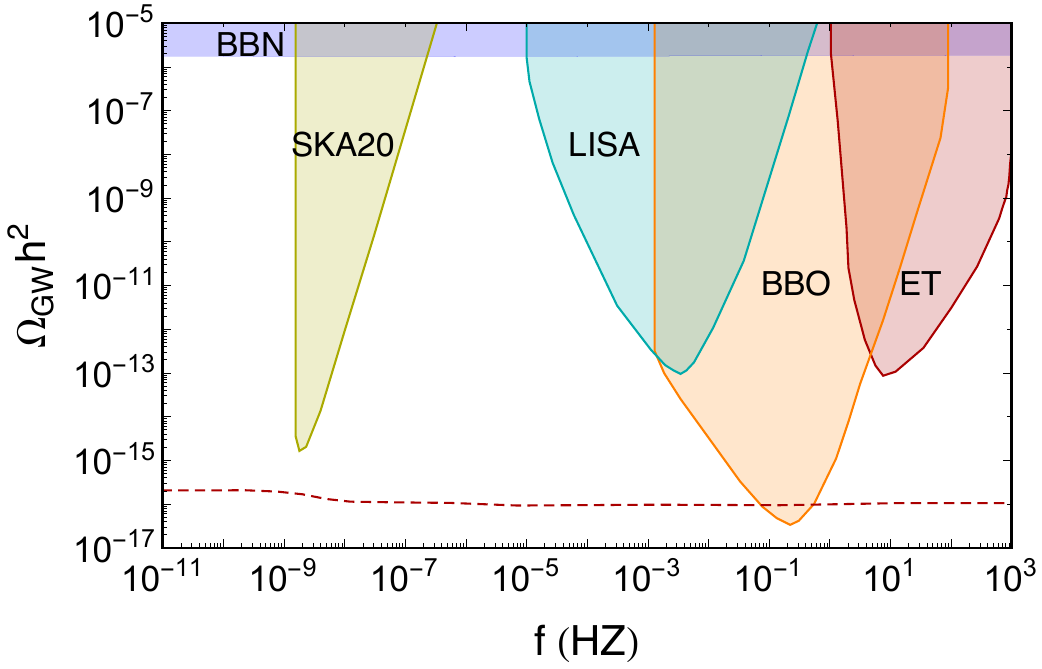}
\caption{Plot of the PGW spectrum versus $f$, for $n_T=0$ and $A_S\sim10^{-9}$~\cite{Planck:2018jri}. The colored regions correspond to the projected sensitivities of several GW observatories~\cite{Breitbach:2018ddu}.}
\label{Fig5}
\end{figure}

The scale dependence of the tensor power spectrum is parameterized by $\mathcal{P}_T(k) = A_T\left(\dfrac{k}{\tilde k}\right)^{n_T}$,
where $n_T$ is the tensor spectral index and $\tilde k=0.05\,\mathrm{Mpc}^{-1}$ a characteristic wave number scale. The amplitude $A_T$ of tensor perturbations is related to that of scalar perturbations by $A_T=r A_S$, where $r\simeq0.06$ is the tensor-to-scalar ratio. 

The PGW relic density~\eqref{Spt} is plotted versus the frequency $f$ in Fig.~\ref{Fig5} (red dashed curve). Physical units have been restored for direct comparison with literature.  The projected sensitivity has been displayed for several GW observatories, including the LISA interferometer, the Einstein Telescope (ET), the Big Bang Observer (BBO) and the Square Kilometre Array (SKA) telescope~\cite{Breitbach:2018ddu}. Additionally, the Big Bang Nucleosynthesis (BBN) bound follows from the constraint on the effective number of neutrinos.

\subsection{GUP-based cosmology}
Let us extend the above analysis to the GUP framework. Using the Friedmann equation~\eqref{H} and the spectrum~\eqref{Ttps}, the GUP-modified relic density takes the form
\begin{eqnarray}
 \Omega_{\mathrm{GW}}(\tau,k)\!&\simeq& \!\left[\frac{a_{\mathrm{hc}}}{a(\tau)}\right]^4\left[\frac{H_{\mathrm{hc}}}{H_{\mathrm{GR}}(\tau)}\right]^2\left[\frac{H_{\mathrm{GR}}(\tau)}{H(\tau)}\right]^2\frac{\mathcal{P}_T(k)}{24} \nonumber \\[1mm]
&&\hspace{-2.1cm}=\Omega^{\mathrm{GR}}_{\mathrm{GW}}(\tau,k)\left[\frac{H_{\mathrm{GR}}(\tau)}{H(\tau)}\right]^2\left[ \frac{a_{\mathrm{hc}}}{a_{\mathrm{hc}}^{\mathrm{GR}}}\right]^4\nonumber \left[ \frac{a^{\mathrm{GR}}(\tau)}{a(\tau)}\right]^4\left[ \frac{H_{\mathrm{hc}}}{H_{\mathrm{hc}}^{\mathrm{GR}}}\right]^2   \,, 
\\
\label{eq:PGWBar0}
\end{eqnarray}
where $H(\tau)$, $a(\tau)$ are the GUP-corrected Hubble rate and scale factor, respectively, while the subscript ``GR'' labels the corresponding standard quantities
in GR. 

By further implementing the flatness condition $H(z=0)/H_0=1$, we finally acquire
\begin{equation}
 \Omega_{\mathrm{GW}}(\tau_0,k)
\ \simeq \ \Omega^{\mathrm{GR}}_{\mathrm{GW}}(\tau_0,k)\left[ \frac{a_{\mathrm{hc}}}{a_{\mathrm{hc}}^{\mathrm{GR}}}\right]^4\left[ \frac{H_{\mathrm{hc}}}{H_{\mathrm{hc}}^{\mathrm{GR}}}\right]^2   \,, \label{eq:PGWBar} 
\end{equation}
which provides us with the GUP-modified expression of the GW spectrum. Note that corrections are embedded in the factors $H_{hc}$ and $a_{hc}$, respectively.

\begin{figure}[t]
\centering
\includegraphics[width=8.5cm]{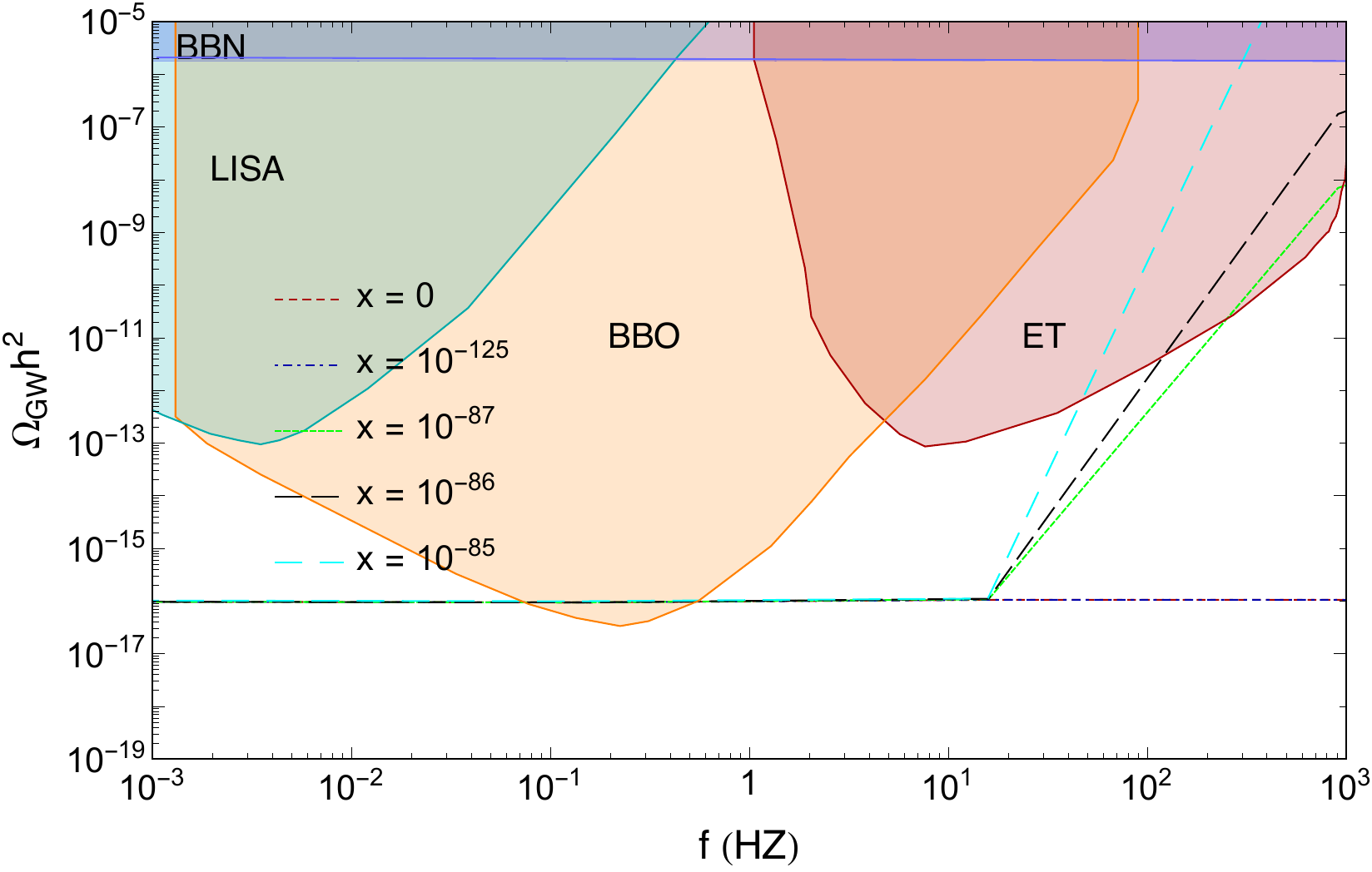}
\caption{Plot of the modified PGW spectrum versus $f$. We restricted to the frequency range $[10^{-3},10^3]\,\mathrm{Hz}$.}
\label{Fig6}
\end{figure}

The PGW profile~\eqref{eq:PGWBar} is displayed as a function of $f$ in Fig.~\ref{Fig6}. We restricted to the frequency domain $[10^{-3},10^3],\mathrm{Hz}$, which appears more sensitive to GUP effects. \textcolor{black}{In this analysis, we assume that quantum gravity influences cosmic evolution primarily at the background level. In other words, all potential corrections induced by the GUP are encapsulated within the modified expression for the Hubble parameter. This assumption remains valid for small deviations from GR, which is the specific regime under investigation in this work\footnote{\textcolor{black}{More broadly, quantum gravity effects could also manifest at the level of linear perturbations, potentially altering the transfer functions and power spectra of both scalar and tensor perturbations generated during the inflationary epoch~\cite{Lewis:1999bs, DAgostino:2023tgm}. However, the exploration of such effects is beyond the scope of this study and is deferred to future investigations.}}. Consequently, since the tensor power spectrum, Eq.~\eqref{TPS}, is strongly dependent on the Hubble parameter, the modifications introduced by the GUP lead to non-negligible corrections to the relic density in Eq.~\eqref{Spt}.  Notably, as GUP-induced corrections to $H$ become increasingly significant at higher redshifts (see Fig.~\ref{Fig2}), or equivalently at higher energy/momentum scales, the most substantial deviations in $\Omega_{GW}h^2$ emerge at frequencies exceeding a certain threshold, in line with Eq.~\eqref{freq}. In other words, this establishes the existence of a threshold frequency, above which the corrections become pronounced, while below this frequency, our extended model remains nearly indistinguishable from standard cosmology.  It is worth noting that a similar enhancement in the PGW spectrum at high frequencies has been observed in modified cosmological frameworks, including scalar-tensor theories and extra-dimensional gravity (see the discussion in the concluding section). Furthermore, the seemingly sudden variation in the value of the relic density is intrinsically tied to the specific scaling adopted for the parameter $\beta$ (or equivalently, $x$).
As depicted in Fig.~\ref{Fig6}, for the sake of graphical clarity, we let \( x \) vary over 40 orders of magnitude, transitioning from $x \simeq 10^{-125}$ (where the model is indistinguishable from GR) to $x \simeq 10^{-87}$. In principle, the variation of the GUP parameter could be implemented more gradually, as done in Fig.~\ref{Fig2}. However, such a gradual approach would render it significantly more challenging to visually discern the effects arising from GUP corrections.
}

Although deviations from GR are hardly detectable for $\beta\sim\mathcal{O}(1)$, i.e., $x\sim\mathcal{O}(10^{-125})$ (the blue dot-dashed and red dashsed curves are nearly indistinguishable), GUP signatures could still be observed by the upcoming ET, should $\beta$ be larger (green and black dashed curves). Such a scenario is not atypical in cosmological contexts, where greater values of $\beta$   are not excluded a priori (see~\cite{Giardino:2020myz,Bosso:2023aht}). 

Notably, by ensuring that the BBN constraint is not violated in the considered frequency range, we find  $x<\mathcal{O}(10^{-85})$ (cyan dashed curve), which in turn sets 
\be
\label{fbound}
\beta<\mathcal{O}(10^{39})\,.
\ee 

To the best of our knowledge, this is one of the most stringent constraints of cosmological/astrophysical origin\footnote{Several attempts to constrain $\beta$ have been conducted in low-energy quantum theory~\cite{Giardino:2020myz,Bosso:2023aht}. While the resulting bounds are among the most stringent, caution is warranted when contrasting predictions from different frameworks, as the GUP may influence systems in disparate ways that are subject to diverse conceptual interpretations~\cite{Bosso:2023aht}.  
Therefore, for our comparative analysis, it makes sense to focus on constraints derived from cosmological and astrophysical contexts that are similar to the one being discussed in this paper.}.  
For comparison, we remind that the analysis of black hole shadow sets $\beta<10^{90}$~\cite{Neves:2019lio}, to contrast with $\beta<10^{69}$ inferred from measurements of the Perihelion precession from Solar system data~\cite{Scardigli:2014qka}. Such constraints are improved to $\beta<10^{59}$ using full data cosmology~\cite{Giardino:2020myz}. Tighter bounds are apparently obtained in approaches that violate the Equivalence Principle~\cite{Ghosh:2013qra,Gao:2017zch}, although this violation may entail serious drawbacks~\cite{Casadio:2020rsj}. 
With specific reference to GWs, the study of modified dispersion relations for gravitons implies $\beta<10^{60}$~\cite{Feng:2016tyt}, while the observation of GW signals through resonant bar detectors yields the more restrictive condition $\beta<10^{33}$~\cite{Marin:2013pga}. On the other hand, any direct comparison with~\cite{Das:2022hjp} might be conceptually challenging, due to the analysis in that context being framed within modified gravity theories (the interested reader may refer to~\cite{Giardino:2020myz,Bosso:2023aht} for a more exhaustive review).

Therefore, the result~\eqref{fbound} underscores the potential role of high-precision cosmology in the framework of QG phenomenology,  highlighting the importance of GWs in advancing our understanding of QG.

\section{Conclusions and Outlook}
\label{Co}

The Generalized Uncertainty Principle is a one-parameter deformation of the Heisenberg relation,  arising from phenomenological considerations in QG theories and gedanken experiments on micro black holes. Inspired by the GUP deformation of the entropy area law, we analyzed cosmological equations for the flat FRW Universe within Jacobson's approach.
Interestingly, we found that the GUP correction acts as an effective quintessence-like DE, $\rho_{DE}\sim \beta H^4$, supplementing the cosmological constant and influencing the early evolution of the Universe, while aligning with the $\Lambda$CDM model at present time. 

We applied this extended model to study the growth of matter perturbations in the early stages of the Universe. Employing the Top-Hat SC formalism, we  extracted the dynamical equation for density fluctuations. We observed that the GUP tends to suppress the gravitational evolution of the matter density contrast, leading to a delayed formation of large-scale structures. 

As a further study, we addressed implications for the relic density of PGWs, showing that the GW spectrum is enhanced at higher frequency as the GUP parameter $\beta$ increases.  Specifically, if any deviation from GR is detected with the experimental sensitivity of the upcoming ET, it could be interpreted as a potential QG signature.
\textcolor{black}{It is interesting to note that a similar high-frequency boost in the PGW profile has been achieved in modified cosmological models based on scalar-tensor and extra-dimensional gravity~\cite{Bernal:2020ywq}. 
By parametrizing the Hubble rate as $H(T)=A(T) H_{GR}(T)$, where $A(T)=1+\eta\left(T/T_*\right)^\nu$ and $\nu$, $\eta$ are dimensionless parameters, it has been shown that the PGW spectrum is blue-tilted with respect to the GR curve for $\nu>0$. As a future perspective, it would be compelling to delve deeper into this common prediction, possibly uncovering a connection between these modified gravity models and the GUP at a more fundamental level. }
Additionally, on a more quantitative level, we found that the BBN bound excludes values of $\beta$ higher than or equal to $10^{39}$, providing one of the most stringent astrophysical/cosmological constraints.

Some other aspects are yet to be investigated: first, we aim to further explore the implications of GUP-based cosmology in the very early Universe. In this respect, inflation provides an insightful window into QG, offering a natural framework to test extensions of GR. Moreover, the epoch between the end of inflation and the beginning of the radiation-dominated era is mostly uncharted territory, with several UV-complete scenarios proposing non-standard cosmological theories or modifications to gravity. From this perspective, the timing of our study is particularly relevant, given that upcoming generations of GW observatories will span a wide range of frequencies and amplitudes, potentially revealing new physics related to the era before BBN. 
Work along these directions is already underway and will be elaborated elsewhere.

\bibliography{Bib}

\end{document}